\documentclass[fleqn,usenatbib]{mnras}

\usepackage{newtxtext,newtxmath}

\usepackage[T1]{fontenc}
\usepackage{ae,aecompl}

\usepackage{graphicx}	
\usepackage{amsmath}	
\usepackage{amssymb}	
\usepackage{gensymb}
\usepackage{verbatim }
\usepackage{float}
\usepackage{array}
\usepackage{breqn}

\title[Time Evolution Of Cyclotron Line of Her X-1 ]{Time Evolution Of Cyclotron Line of Her X-1; A Detailed Statistical Analysis Including New ASTROSAT Data }

\author[Bala et al.]{
S. Bala$^{1}$\thanks{E-mail: bala@iucaa.in},
D. Bhattacharya$^{1}$,
R. Staubert$^{2}$,
C. Maitra$^{3}$,
\\
$^{1}$Inter University Centre for Astronomy and Astrophysics, Ganeshkhind, Post Bag 4, Pune 411007, India. \\
$^{2}$ Institut f\"ur Astronomie und Astrophysik, Kepler Center for Astro and Particle Physics, Sand 1, D-72076 T\"ubingen, Germany\\
$^{3}$Max-Planck-Institut fur extraterrestrische Physik, Gießenbachstraße, 85748 Garching, Germany \\
}
\date{Accepted XXX. Received YYY; in original form ZZZ}

\pubyear{2020}

\begin{document}

\label{firstpage}
\pagerange{\pageref{firstpage}--\pageref{lastpage}}
\maketitle
%
\begin{abstract}
 The cyclotron line feature in the X-ray spectrum of the accretion powered pulsar Her X-1 has been observed and monitored for over three decades. 
The line energy exhibited a slow secular decline over the period 1995-2014, with a possible (not confirmed) indication of a reversal thereafter.  Recent works have shown that the temporal evolution of the line energy may be modelled as a flattening after an earlier decrease until MJD 55400 ($\pm200)$. 
In this work, we present the results of ASTROSAT observations in the context of earlier data and offer a common interpretation through a detailed study of temporal and flux dependence. 
We find that the variation of the line energy does not support an upward trend but is consistent with the reported flattening after an earlier decrease until MJD $54487^{+515}_{-469}$.
  
\end{abstract}




\section{Introduction}
Cyclotron Resonance Scattering Features (CRSF), also referred to as Cyclotron Lines, 
constitute an important tool for measuring the magnetic field strength of accreting
neutron stars in X-ray binary systems.  A long-term study of the cyclotron
line can reveal the evolution of the magnetic field in the x-ray emitting region.
In the highly magnetized plasma accumulated on the neutron star polar caps, the cyclotron
lines are produced by resonant scattering of photons off electrons
moving perpendicular to the magnetic field, and appear
as absorption features against the X-ray continuum. The line
energy ($E_{\text{cyc}}$) is related to the magnetic field as $B_{12}=(1+z)E _{\text{cyc}}/11.6$
keV, where $B_{12}$ is the field strength in units of $10^{12}$
Gauss, and $z$ is the gravitational red shift.\par

Her X-1 is a 1.24 s period accretion-powered X-ray pulsar in a 1.7 day circular
orbit with a normal stellar companion, HZ Her. The edge-on view $(i>80\degree)$ of the accretion 
flow in this system reveals a rich variety of temporal and spectral features, which have been the subject
of concerted study over many years. In addition to the basic spin and orbital periodicity,
this binary system has been known to display a 35 day long cycle which consists of a "Main-On" (high X-ray flux) and a "Short-On" states, separated by two "Off" states \citep{1976ApJ...209L.131J,2013A&A...550A.110S}. The sharp increase of the flux ("turn on") indicates the beginning of a cycle. This generally takes place either around binary phase 0.2 or 0.7.
The cyclotron line in Her X-1 or in an X-ray binary was first detected in a 1976 balloon observation (\citet{1977NYASA.302..538T}, 
\citet{1978ApJ...219L.105T}). At present nearly 36 sources are known to exhibit CRSF in their X-ray spectra \citep{2019A&A...622A..61S}.
The source Her X-1 is known to exhibit a positive correlation between  $E_{\text{cyc}}$ and the X-ray luminosity $L_{x}$\citep{2007ESASP.622..465S}. A long-term decay of  $E_{\text{cyc}}$ in Her X-1 was first discovered by \citet{2014arXiv1412.8067S}. Before 1991 the mean cyclotron line energy was around $\sim$35 keV (derived using HEAO-A4, Mir-HEXE, and
Ginga observations \citep{1990ApJ...348..641S}).  Subsequently a very high value of nearly 44 keV (mean of the values observed in
1993 and 1996) was noticed from CGRO/BATSE data \citep{1996AIPC..384..172F}. There was no observation that 
captured the intervening rise. Later, observations with BeppoSax and RXTE in 1997
found the line energy to be around 40-41 keV \citep{2001ApJ.Gruber.562..499G}. A regular monitoring was initiated at this stage, and 
the accumulated data since \ensuremath{\sim} 1996 appeared to show that  $E_{\text{cyc}}$ has both 
luminosity and time dependence \citep{2014A&A...572A.119S}, with a slow secular decay of line 
energy with time.
Further observations with INTEGRAL, Suzaku and NuSTAR fit well with a model assuming
a continued decay of  $E_{\text{cyc}}$ until 2015, with a decay rate of $0.260\pm0.14$
keV/yr \citep{2016A&A...590A..91S}. However in
2016 NuSTAR and INTEGRAL observations of Her X-1 began to indicate a possible upward trend in line energy \citep{2017A&A...606L..13S} (which is not confirmed so far), possibly similar to the jump observed between 1991 and 1993.
A recent work with SWIFT/BAT data claims that  $E_{\text{cyc}}$  may in fact have been constant since 2012, 
with an attendant change in its dependence on luminosity \citep{2019MNRAS.484.3797J, 2019JHEAp..23...29X}.\par 
In this work we investigate both the luminosity and the temporal dependence of  $E_{\text{cyc}}$ of Her X-1, 
using  all previous data and new observations from the Indian AstroSat mission.  In section 2 we introduce AstroSat and its instruments, as well as the standard data reduction tools. In section 3 we give an account of the method we have followed to analyze the data. In section 4 we compare different models of  $E_{\text{cyc}}$ evolution. 
In section 5 we present our results and discuss possible physical reasons behind the observed behaviour of cyclotron line energy. Errors, wherever quoted, correspond to 68\% confidence levels, unless otherwise mentioned.

 \begin{figure*}
\centering
\includegraphics[width=17cm]{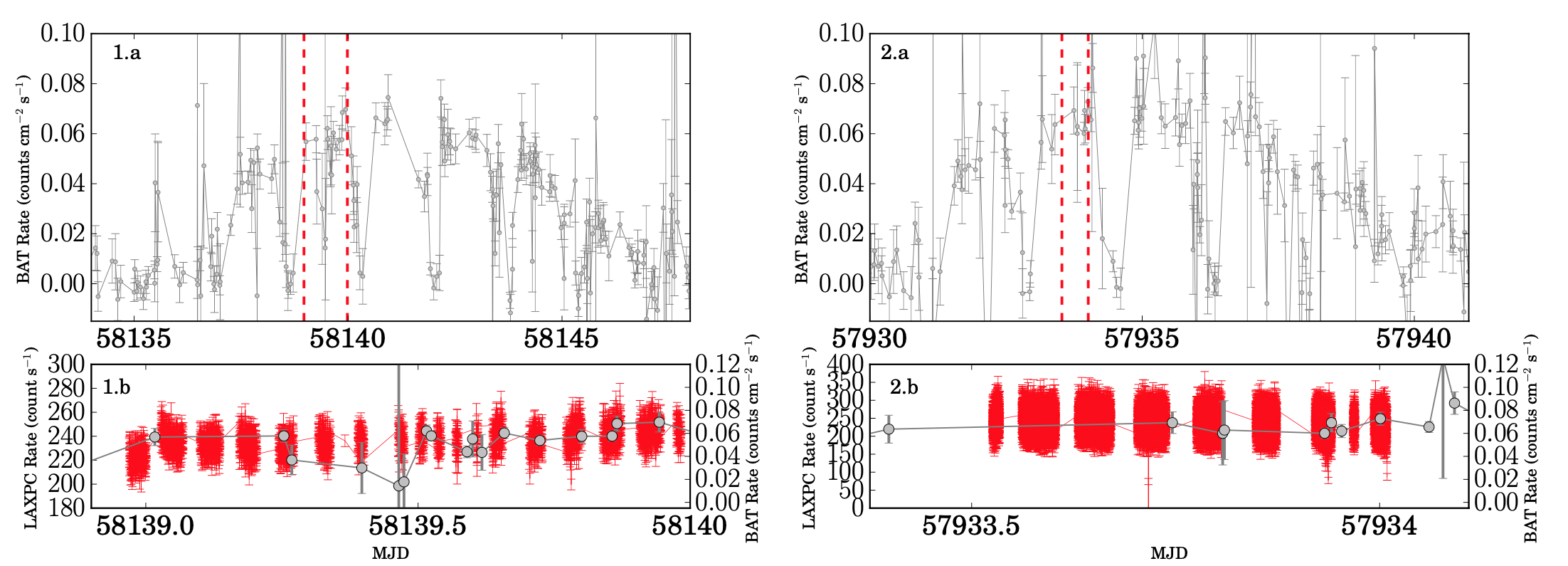}
\caption{Her X-1 light curve, upper panels (1.a,2.a) are the \textit{Swift}/BAT light curve (gray points) around our observations. The duration of our observations are shown by the vertical red dashed lines. The Lower panels show the zoom in plot during our observations. The 15-50 keV LAXPC10 light curves (red) of dataset-1 (1.b) and dataset-2 (2.b) (red) over plotted with the BAT light curve (gray circles).}
\label{fig:all_fig}
\end{figure*}

\begin{table}
\caption{Observation log, errors are mentioned with 1$\sigma$ (68\%) confidence range.}
\centering
\begin{tabular}{p{0.8cm}p{0.6cm}p{0.7cm}p{1.6cm}p{1.3cm}p{1.cm}}
\hline
Obs & Center  & Duration & BAT count  & ASM flux  & $E_{\text{cyc}}$ \\
Date & MJD & (ks) & (cm$^{-2}$s$^{-1}$) & (counts/s) & (keV) \\
\hline \hline
21 Jan 2018 & 58141 & 59.6 & $0.061\pm0.002$ & $5.67\pm0.30$ & $36.89^{+0.49}_{-0.44}$ \\
\hline
30 Jun 2017 & 57934 & 35.4 & $0.07\pm0.002$ & $6.51\pm0.30$ & $37.73_{-0.58}^{+0.65}$ \\
\hline \hline
\label{observation_log}
\end{tabular}
\end{table}
\section{Observations and Data Reduction}
 In this work we use two datasets obtained using AstroSat's \citep{2014_sing_AstroSat} Large Area X-ray Proportional Counters (LAXPC) \citep{Agrawal2017} and the Soft X-ray Telescope (SXT) \citep{Singh2017}. AstroSat is India's first multi-wavelength satellite, carrying detectors in the visible, near and far ultraviolet (NUV/FUV) as well as soft, medium and hard X-ray bands. AstroSat consists of five principal scientific instruments, (i) a Soft X-ray Telescope (SXT) (0.3-8.0 keV), (ii) three Large Area X-ray Proportional Counters (LAXPCs) (3.0-80.0 keV), (iii) a Cadmium-Zinc-Telluride Imager (CZTI) (25-200 keV), (iv) two Ultra-Violet Imaging Telescopes (UVITs), capable of observing the sky in the Visible, Near Ultraviolet and Far Ultraviolet regions of the electromagnetic spectrum and (v) three Scanning Sky Monitors (SSMs), to scan the sky for long term monitoring of bright X-ray sources and the detection of bright X-ray transients. AstroSat also carries on board a Charged Particle Monitor (CPM).\par
 
The first dataset in Table--\ref{observation_log} was obtained from an observation proposed by us (ObsID 9000001854, PI: Suman Bala) and carried out over 20--22 Jan 2018, during the \textit{Main-On} phase, avoiding eclipses. The second dataset (Obs ID: 9000001348, PI: Denis Leahy) carried out between 29-30 June 2017, was during a \textit{Main-On} phase but also contains \textit{Main-on dips}.
 Data points from the \textit{Main-on dips} have been removed from our spectrum analysis.  The datasets are retrieved from the AstroSat open public archive\footnote{\href{https://astrobrowse.issdc.gov.in/astro\_archive/archive/Home.jsp}{https://astrobrowse.issdc.gov.in/astro\_archive/archive/Home.jsp}}.
The AstroSat observation epochs are marked on the SWIFT/BAT lightcurve shown in the upper panel of Fig--\ref{fig:all_fig}. BAT and LAXPC (15-50 keV) light curves within the AstroSat observation period are compared in the lower panel of Fig-\ref{fig:all_fig}, showing similar variations.\par 



\subsection{Large Area X-ray Proportional Counters (LAXPC)}
The LAXPC consists of a set of three non-imaging proportional counters (LXP10, LXP20, LXP30) nearly co-aligned with the satellite axis. The LAXPC operates in the energy range 3--80 keV with the energy resolution varying between  10\% to 20\%. A detailed description of the characteristics of the LAXPC instrument can be found in \citet{2017ApJS_antia_LAXPC}.\par

For LAXPC, we use the level 1 data from the archive and the
\texttt{laxpcsoft} package available at the AstroSat Science Support Cell (ASSC)\footnote{\href{http://AstroSat-ssc.iucaa.in}{http://astrosat-ssc.iucaa.in}} for analysis. First, we create a combined event file of all three LAXPC units using the standard procedure \texttt{laxpc\_make\_event}. After that we use \texttt{laxpc\_make\_stdgti} to generate a good time interval (\textit{gti}) file, removing intervals of passage through the South Atlantic Anomaly (SAA) and Earth occultation. Using this \textit{gti} file we create individual light-curves of each LAXPC unit (\texttt{laxpc\_make\_lightcurve}), and check for any anomalies in counts rates. We remove these anomalous intervals, all periods of binary eclipses and the time interval of absorption dips from the \textit{gti} and create a filtered \textit{gti} file for every LAXPC unit, as advised by the LAXPC instrument team. Finally we use the modified \textit{gti} file to create spectra (using \texttt{laxpc\_make\_spectra}) for every individual LAXPC unit. The software also provides an appropriate grouping of the spectrum, oversampling the spectral resolution ($\sim$15\%) of the instrument by at most a factor of 3. The module \texttt{laxpc\_make\_backspectra} is used to generate the corresponding background spectrum,  having gain close to our observation. The LAXPC pipeline creates the Response Matrix File (\texttt{.rsp}) corresponding to each observation and LAXPC unit. We further augment the response file with an additional Auxiliary Response File (ARF) created by us, as described in the Appendix--\ref{ARF}.\par 

At the time of our observations the LAXPC unit LXP30 had a rapidly changing gain due to a gas leak, and LXP20 had an uncertain spectral response caused by gas impurities and incomplete recording of Xenon K escape events.  These two units were thus not suitable for accurate spectroscopic work.  We therefore use data only from LXP10, in the energy range 
5-60 keV since above 60 keV the background dominates and any error in background estimation can 
affect the spectrum significantly. 
Below 5~keV the background and the response of LAXPC are not yet fully understood and can 
leave unmodelled residuals in spectral fitting. For LAXPC10, the 5-60 keV net source 
count rates are $427.3\pm0.3$ counts/s and $448.8\pm0.3$ count/s respectively in the two datasets, and the corresponding 
background counts rates are $166.2\pm0.3$ counts/s and $160.4\pm0.3$ counts/s.\par
\subsection{Soft X-ray imaging Telescope (SXT)}
The SXT \citep{Singh2017} is a focussing telescope with a focal length of 2~m and a thermo-electrically cooled CCD in the focal plane. Its operating energy range is 0.3--7.0 keV with an energy resolution $\sim 150$~eV at 6 keV. The field of view of SXT is of $\sim$40$^{\prime}$ diameter with a Point Spread Function (PSF) of $\sim$2$^{\prime}$
Full Width at Half Maximum (FWHM). The SXT observations used in this work have been carried out in the "Photon Counting" (PC) mode, where the data from the entire CCD of $600\times600$ pixels are collected. Currently the lower energy threshold of SXT is set at 105 eV, $4\sigma$ above the electronic noise peak but the quantum efficiency of the detector is poorly determined below 0.7 keV. So, for our analysis, we have used the SXT Spectrum in the energy range 0.7--7.0 keV.\par

For SXT, we have used the level-2 data products from the archive. The level-2 data, namely the cleaned event files for individual orbits, are created by the SXT Payload Operation Centre (POC) using the \texttt{SXTPIPELINE version-1.4a} that takes care of time tagging of the events,  calibration of the source events, the coordinate transformation from detector to sky coordinates, the pulse height amplitude (PHA) construction for each event, the conversion from the event PHA to pulse invariant (PI), bias subtraction and adjustment, flagging of the bad pixels and removal of hot and flickering pixels.  We use a standard julia script, developed by the SXT instrument team, to create a single event file, by merging the event files of all orbits.  A source region of 10 arcmin radius, which includes 90\% of the source photons, has been selected from the image to generate the light curve and energy spectrum using the XSELECT task of HEASOFT\footnote{\href{https://heasarc.gsfc.nasa.gov/docs/software/heasoft/}{https://heasarc.gsfc.nasa.gov/docs/software/heasoft/}} version 6.25. We used the background file \texttt{SkyBkg\_comb\_EL3p5\_Cl\_Rd16p0\_v01.pha} available at the ASSC. The updated response Matrix \texttt{sxt\_pc\_mat\_238new.rmf} and Auxiliary Response File (ARF)  \texttt{sxt\_pc\_excl00\_v04\_mJ\_final\_scl.arf} have been provided by G. C. Dewangan (private communication). The net SXT count rate for the dataset-1 and dataset-2 have been found to be $19.72\pm0.04$ and $22.06\pm0.05$ counts/s respectively, much less than the threshold ($\sim$ 40 counts/s) above which the photon pile-up takes place. The SXT spectra have been binned considering at least 30 counts per energy bin.

\section{Data Analysis} 
The hard X-ray continuum produced in the accretion column so far does not have any generally accepted physical model. In some special cases (e.g., for 4U 0115+63; \citet{2009A&A.Ferrigno.498..825F}), models based on the work by \citet{2007ApJ.Becker.Wolff.654..435B} have been used to fit the continuum.  Using the model of \citet{2007ApJ.Becker.Wolff.654..435B}, \citet{2016ApJ...831..194W} has shown that the spectrum of Her X-1 obtained from a 2012 NuSTAR observation can be explained as continuum produced by a radiation dominated shock. Still, the model of \citet{2007ApJ.Becker.Wolff.654..435B} has not been universally applied.
It is therefore a common practice to use empirical models, like power law with an exponential roll over, to model the spectral continuum of X-ray pulsars. The continuum of Her X-1 is best described by a phenomenological power law with high-energy cutoff (\texttt{highecut}). Previous results of the long term evolution of the cyclotron line energy of Her X-1 have been reported using this continuum model. So, in this work we use the same model to fit the continuum for consistency.\par

The spectral analysis is done using \texttt{XSPEC} version: 12.10.0. From both the datasets, we have used the SXT data in the energy range of 0.7-7.0 keV and the LAXPC data in the range 5.0-60.0 keV. We have included a 2\% systematic error in the SXT data and 1\% in LAXPC data for the combined fitting (as 
recommended by the instrument teams). We have used the parameter \texttt{constant} to account for cross-instrument calibration 
uncertainties. The constant factor for LAXPC is frozen to 1.0 and for the SXT it is kept free. For SXT the gain adjustment with the
slope fixed at 1 and an offset, has been kept free. A black body component ($kT\sim$0.1 keV) and a Gaussian Iron line at 
$\sim{6.5}\pm0.18$ keV have been used to model the combined spectrum.  As the width of the iron line could not be constrained
well, it has been frozen to the previously reported value of 0.22 keV from NuSTAR observations \citep{2013arXiv1309.5361F}. A 
recent study of Her X-1 has shown the presence of an emission line at $\sim0.94$ keV \citep{2019ApJ...871..152L}. Thus, we have
included a Gaussian emission line around 0.94 keV in our combined fitting. The galactic absorption for Her X-1 is always found to be very low \citep{2013arXiv1309.5361F, 2019ApJ...871..152L}, 
whereas including a partial-covering absorber \texttt{tbpcfabs} 
\citep{2000ApJ...542..914W} improves the fit.  The 
\texttt{highecut} model shows a sharp feature at the cutoff 
energy (e.g. \citet{1997ESASP.Kretschmar.382..141K}, 
\citet{1999A&A.Kreykenbohm..341..141K}), and a Gaussian optical-depth 
 profile (\texttt{gabs}) is often used to smooth it  (e.g. \citet{2002ApJ...580..394C, 2013arXiv1309.5361F}). So, we have included a \texttt{gabs} with its energy tied to the cutoff energy, keeping the width $\sigma_c$ and depth $\tau_c$ free. A 
 cyclotron line feature is generally modeled with either a \texttt{cyclabs} or a \texttt{gabs} profile.  Following earlier works, we have used a Gaussian optical-depth (\texttt{gabs}) to model the cyclotron line.  The complete model used for the combined fit of LAXPC and SXT is,
 
\begin{equation}
\begin{aligned}
\text{I}_{\text{c}}= &\text{ Constant*tbpcfabs*(Gauss+Gauss+bbody+gabs*gabs}\\
&\text{*highecut*powerlaw)} 
\end{aligned}
\label{model}
\end{equation}

 After fitting the LAXPC spectrum, we have observed some overall unmodelled residuals. This could be either due to a poor response function or inadequate background modelling. So, we fixed the continuum parameters at values derived after studying NuSTAR observations of Her X-1 (\citet{staubert_2020_AA}, submitted to A\&A). The continuum parameters corresponding to the observed flux during our observations are found to be,  $E_{\text{cut}}$= 20.37 keV and 20.62 keV,  $E_{\text{fold}}$= 10.1 keV and 10.15 keV and Photon Index = 0.99,0.99 for dataset-1 and dataset-2 respectively. At high energies the spectrum is found to deviate from the model, mostly due to improper modelling of LAXPC background spectra. To counter that we have used the original
 background spectra generated by LAXPC pipeline with an adjustable rescaling factor. The best fit is achieved when the original background spectra are rescaled by factors 1.05 and 1.04 for dataset-1 and dataset-2 respectively. We also find that at high energies the original background spectra deviate from the spectra obtained while Her X-1 was occulted by the Earth,
 and the same rescaling factors as above resolved those differences (Appendix--\ref{back_earth_occ}). Even after this, we find some residuals in the overall LAXPC spectra, which are also observed in the LAXPC spectra of other strong sources. This systematic deviation may be attributed to the uncertainties in the response of LAXPC. So, we have  created an Auxiliary Response (ARF), using the Crab spectrum, in an attempt to remove these artifacts. The details of the ARF have been given in Appendix--\ref{ARF}. The ARF for dataset-1 has been created using a Crab spectrum observed on 16th Jan, 2018 (MJD 58134, 7 days before our observation) but for dataset-2 we did not find any near-by Crab observation so a corresponding ARF could not be created.

\begin{center}
\begin{table}
\caption{The best fit model parameters of the combined fit of LAXPC and SXT, with and without including our arf. All the uncertainties are measured in 90\% confidence level.}
\renewcommand{\arraystretch}{1.60}
\begin{tabular}{|p{2.0cm}|p{1.4cm}p{1.4cm}|p{1.4cm}|}
\hline
\multicolumn{1}{|c|}{} & \multicolumn{2}{c|}{Dataset-1} & \multicolumn{1}{c|}{Dataset-2} \\
\hline
Parameter & With & Without &  Without \\
        & ARF    & ARF     &  ARF \\
\hline
$N_{\text{H}}$ ($10^{22}$cm$^{-2}$) &    $10.98^{+4.91}_{-4.50}$ & $14.95^{+4.80}_{-4.28}$ &   $19.66^{+3.54}_{-3.54}$  \\
$f$ &                                    $0.12^{+0.03}_{-0.03}$ &  $0.14^{+0.03}_{-0.03}$ &   $0.18^{+0.02}_{-0.02}$   \\
$\Gamma$ &                               $0.99^*$ &  $0.99^*$ &  $0.99^*$ \\
$E_{\text{cut}}$(keV) &                  $20.37^*$ & $20.37^*$ &  $20.62^*$ \\
 $E_{\text{fold}}$ (keV) &              $10.01^*$ & $10.01^*$ & $10.15^*$\\
$kT _{\text{BB}}$ (keV) &               $0.95^{+0.04}_{-0.04}$ & $0.09^{+0.04}_{-0.04}$ &  $0.06^{+0.09}_{-0.05}$ \\
$E(K_{\alpha})$ (keV) &                 $6.53^{+0.11}_{-0.11}$ & $6.59^{+0.11}_{-0.11}$ &  $6.52^{+0.08}_{-0.08}$ \\
$\sigma(K_{\alpha})$ (keV) &            $0.22^*$  & $0.22^*$  &  $0.22^*$  \\
$E _{\text{gauss}}$  (keV) &            $0.93^{+0.05}_{-0.09}$ & $0.93^{+0.05}_{-0.09}$ & $0.89^{+0.07}_{-0.05}$  \\
$\sigma _{\text{gauss}}$  (keV) &          $0.17^{+0.04}_{-0.04}$ &  $0.17^{+0.04}_{-0.04}$ & $0.19^{+0.02}_{-0.06}$  \\
 $E_{\text{cyc}}$ (keV) &                $36.89^{+0.84}_{-0.71}$ & $36.52^{+0.79}_{-0.69}$ &  $37.73^{+1.12}_{-0.92}$ \\
$\sigma _{\text{cyc}}$ (keV) &           $3.48^{+0.66}_{-0.6}$ &  $3.19^{+0.60}_{-0.55}$ & $5.11^{+0.73}_{-0.63}$  \\
$d _{\text{cyc}}$ &                     $5.63^{+0.90}_{-0.77}$ &  $5.12^{+0.82}_{-0.71}$ &  $7.46^{+1.23}_{-1.0}$\\
\hline
$\chi^2$/d.o.f &                        $542.9/529$ & $550.5/529$ & $679.9/598$ \\
\hline \hline

\end{tabular}

Optical depth, $\tau=d/\sqrt{\sigma2\pi}$.  \\
*The parameter value is fixed. 

\label{parameters}
\end{table}
\end{center}


\begin{figure*}
\centering
\includegraphics[width=17cm]{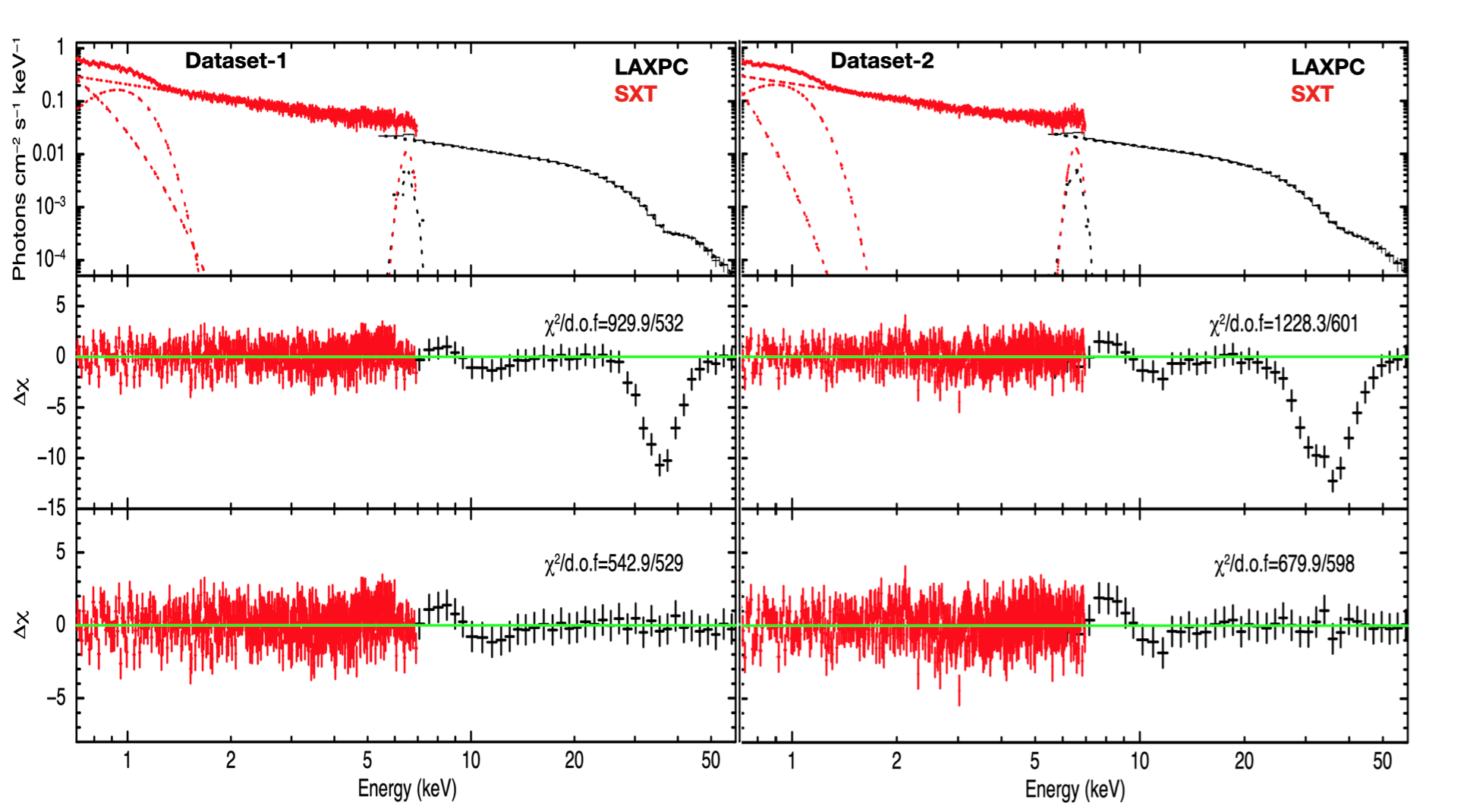}
\caption{Spectrum of Her X-1 observed by AstroSat. In the figure, red represents the SXT data and black represents the LAXPC data. Panels on the left correspond to dataset-1 and those on the right to dataset-2. The top panel shows the unfolded spectrum with different model components. The cyclotron absorption line is shown after fitting a gabs and then setting the depth to zero (middle panels). The lower panels show the best fit delchi plots.}
\label{fig:her_x1}
\end{figure*}

For dataset-1, the cyclotron line energy is found to be $36.89^{+0.84}_{-0.71}$  keV, with a width of $3.48^{+0.66}_{-0.6}$ keV. The fit is found to improve significantly when the cyclotron feature is included, $\chi^2$/d.o.f=542.9/529 compared to that without ($\chi^2$/d.o.f=929.9/532). The best fit parameter values are given in Table--\ref{parameters}. The table displays parameters both with and without the inclusion of the LAXPC ARF derived by us. For dataset-2 the line energy is found to be $37.73^{+1.12}_{-0.92}$ keV and the  $\chi^2$/d.o.f is found to be $1228.3/601$ without the line and $679.9/598$ with the line. In both cases the line has been detected with more than $5\sigma$ confidence level. Fig--\ref{fig:her_x1} shows the unfolded spectra of Her X-1 and best fit delchi plots. We find that our line energies are consistent but the widths are lower than the expected, comparing with other measurements of NuSTAR (\citet{staubert_2020_AA}, submitted to A\&A).  This indicates that further modification in the LAXPC response is required, which is beyond the scope of this work.

\section{Variation of The Cyclotron line energy with luminosity and time}
To examine the dependence of  $E_{\text{cyc}}$ on luminosity and time we carried out a two parameter fit using the well known 
non-linear least squares fitting routine \texttt{mpfit} \citep{2009ASPC..411..251M}. We have also used \texttt{emcee}, (a Goodman and Weare's Affine Invariant Markov chain Monte Carlo 
(MCMC) Ensemble sampler \citep{ForemanMackey2013}) for Bayesian parameter estimation with some suitable flat priors (Table--\ref{Tab:relations}).  In this work, we have used all the data previously 
used in \citet{2014A&A...572A.119S} and \citet{2017A&A...606L..13S} and our recent 
AstroSat results. We use the cyclotron line energies derived without any flux or temporal 
correction. 

To investigate the correlation between the line energy and flux, the \textit{Main-on} flux has been computed by taking the
average count rates of $\sim$3 days around the maximum of the \textit{cleaned} main on phase obtained from \textit{Swift}/BAT light-curve (as suggested in Appendix-A of \citet{2016A&A...590A..91S}).    
We find that the maximum \textit{Main-on flux} corresponding to our AstroSat observations are $0.061\pm0.002$  and $0.07\pm0.002$ counts cm$^{-2}$s$^{-1}$ for dataset-1 and dataset-2 respectively. Using the scaling
factor, [2-10 keV ASM-counts/s]=93.0$\times$[15-50 keV BAT-counts cm$^{-2}$s$^{-1}$] \citep{2016A&A...590A..91S}, we find that the equivalent ASM-counts/s are $5.67\pm0.30$ and $6.51\pm0.30$ respectively.\par

We have described the variation of cyclotron line energy  $E_{\text{cyc}}$ of Her X-1 by Eq--\ref{flux_time_eq}.

\begin{figure*}
\includegraphics[width=17cm]{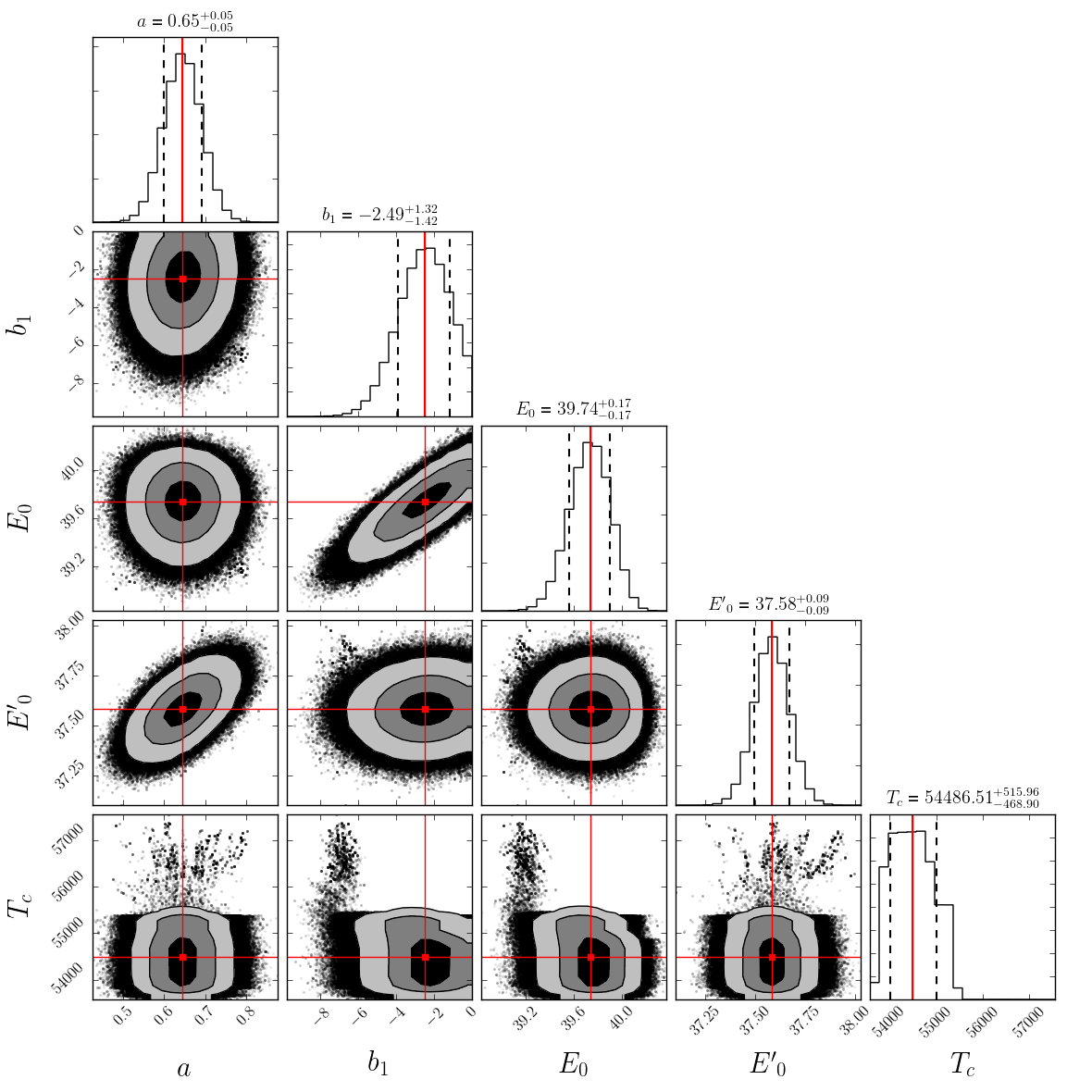}
\caption{The plot shows the 1$\sigma$,2$\sigma$ and 3$\sigma$ parameter contours, and the corresponding marginalized 1D distributions, of 
the five parameters of the model (case--V) in Eq--\ref{flux_time_eq}. The best fit parameter values are denoted by the horizontal and vertical red lines. The median and 1$\sigma$ range are given above the 1D distributions of the respective parameters.}
\label{fig:Best_parameter_distribution}
\end{figure*}

\begin{figure*}
\includegraphics[width=17cm]{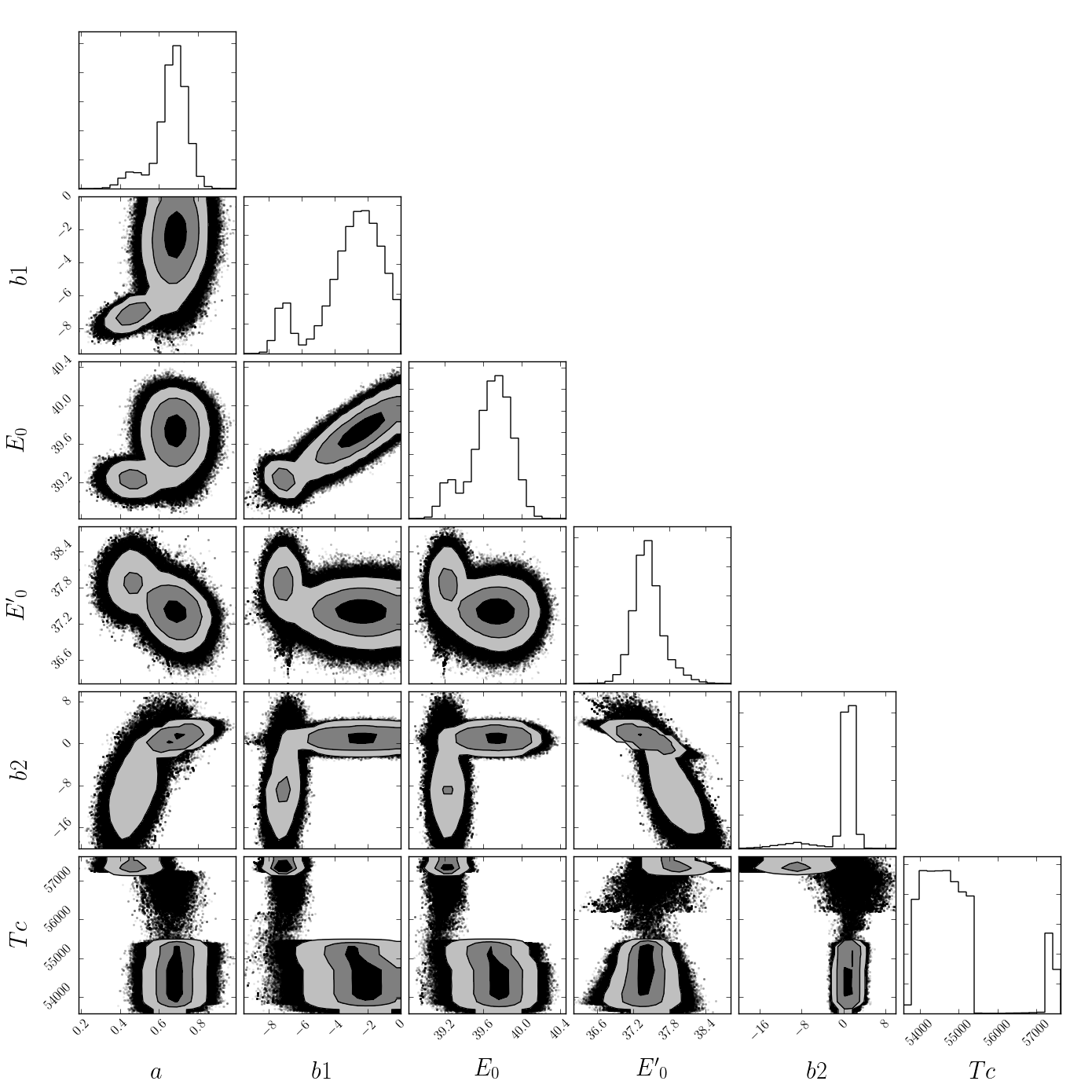}
\caption{The plot shows the 1$\sigma$,2$\sigma$ and 3$\sigma$ parameter contours, and the corresponding marginalized 1D distributions, of case--VII.  Two degenerate solutions can clearly be seen in this figure. }
\label{fig:degenerate_parameter_distribution}
\end{figure*}

\begin{figure*}
\includegraphics[width=17cm]{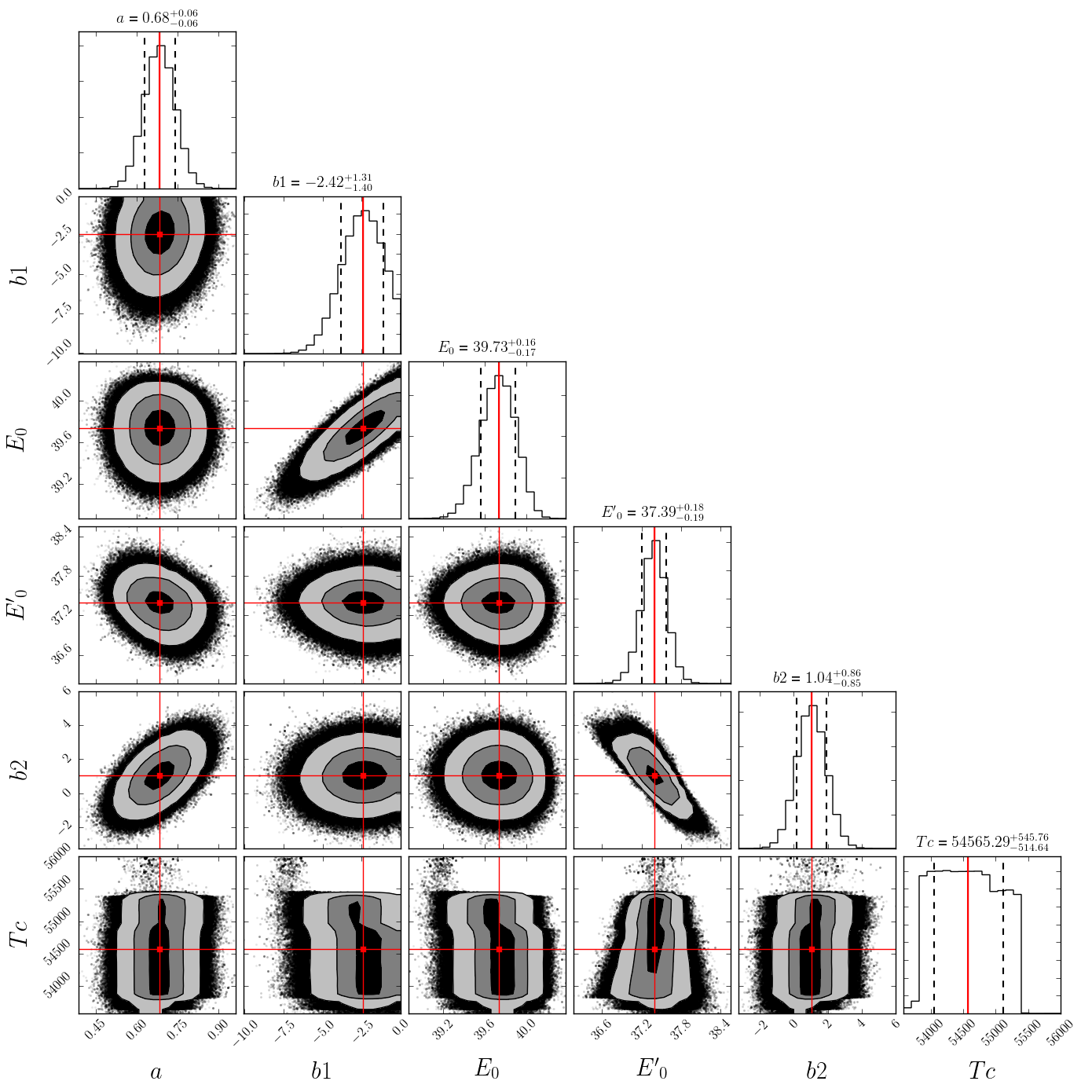}
\caption{The plot shows the 1$\sigma$,2$\sigma$ and 3$\sigma$ parameter contours, and the corresponding marginalized 1D distributions, of case--VII.a. We can clearly see that the degeneracy seen in Fig--\ref{fig:degenerate_parameter_distribution} does not exist any more.  The best fit parameter values are denoted by the horizontal and vertical red lines. The median and 1$\sigma$ range are given above the 1D distributions of the respective parameters.}
\label{fig:Before_56000_parameter_distribution}
\end{figure*}

\begin{figure*}
\includegraphics[width=17cm]{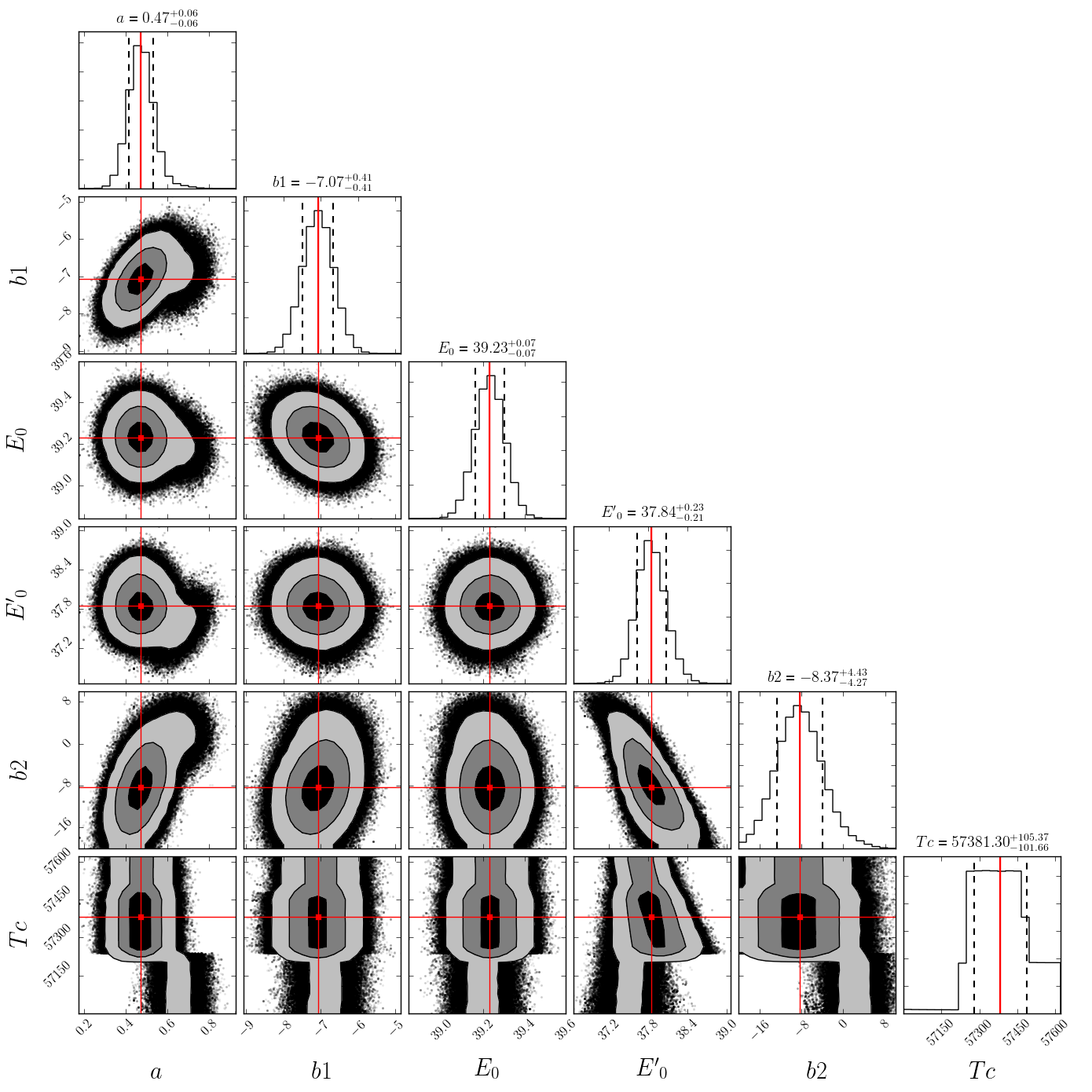}
\caption{Same as Fig--\ref{fig:Before_56000_parameter_distribution}, but for case--VII.b.}
\label{fig:after57000_parameter_distribution}
\end{figure*}

\begin{align}
E_{obs}=& E_{0}+a(F-F_{0})+b_1(T-T_{0}) &  \text{if } 50000 \leq T \leq T_{c} \nonumber\\
=&E'_{0}+a(F-F_{0})+b_2(T-T_{c}) & \text{if } T_c \leq T  \label{flux_time_eq}
\end{align}

Here, $E_{obs}$ is the observed cyclotron line energy at a time $T$ (in MJD) and $F$ is the corresponding ASM \textit{Main-On} flux. $E_0$, $a$, $b_1$, $b_2$, $E'_0$ and $T_c$ are the fit parameters of our models. Here $b_1$ and $b_2$ are in the units of $10^{-4}$ keV/d.
Following earlier works, we set the value of the reference flux $F_0$ to 6.8 ASM-counts/s and the reference time $T_0$ to MJD 53500. We have chosen the parameter ranges in such a way as to reproduce and compare the results from all existing models (e.g. \citet{2014arXiv1412.8067S,2019MNRAS.484.3797J} etc.), intending to find the best description for the variation of the cyclotron line energy of Her X-1. The best fit parameter values for all cases are tabulated in Table--\ref{Tab:relations}.

\textbf{Case--I}: In this case we assume, $30.0<E_0<45$, $0<a<10$,  $b_1=0$,  $E'_0=0$, $b_2=0$, $T_c\to\infty$. This implies that the cyclotron line energy does not depend explicitly on time, and the observed variation in the line energy is only due to the variation of the \textit{Main-On} flux. With this model the best estimated $\chi^2$/d.o.f is found to be 346.7/25. The best fit parameter values are given in Table--\ref{Tab:relations}.

\textbf{Case--II:} Here we assume,  $30.0<E_0<45$, $0<a<10$,  $-20<b_1<0$,  $E'_0=0$, $b_2=0$, $T_c\to\infty$. This implies that the cyclotron line energy has both flux and temporal dependence which is assumed to be the same throughout. This model gives a better fit than case--I, with a $\chi^2$/d.o.f=77.9/24. This suggests the existence of an explicit temporal dependence.  

\textbf{Case--III}: In this case we assume,   $30<E_0<45$, $0<a<10$,  $-20<b_1<0$, $30<E'_0<45$, $b_2=0$, $T_c= 55400$. So, there is both flux and temporal dependence until MJD 55400, after which the time dependence disappears and only the flux dependence remains. This model is equivalent to the model described in \citet{2019MNRAS.484.3797J}. This model gives the $\chi^2$/d.o.f to be 25.8/23.

\textbf{Case--IV}: In this case we assume,   $30<E_0<45$, $0<a<10$,  $-20<b_1<0$, $30<E'_0<45$, $b_2=0$, $T_c= 57341$. This is the same as case--III except that here $T_c$ is set to be MJD 57341 when the line energy is found to be minimum and a 'turn up' is assumed \citep{2017A&A...606L..13S}. The $\chi^2$/d.o.f is found to be 26/23.

\textbf{Case--V}: In this case we assume,   $30<E_0<45$, $0<a<10$, $-20<b_1<0$, $30<E'_0<45$, $b_2=0$, $50000<T_c<59000$. The difference between this and cases III and IV is that here the vale of $T_c$ is found by fitting. We find that this is the best model ($\chi^2$/d.o.f=15.7/22) to describe the observed variation of cyclotron line energy in Her X-1. We find a fitted value of $T_c$ (=$54487^{+515}_{-469}$), close to the previously suggested value $55400\pm200$, by \citet{2019MNRAS.484.3797J}. The distribution of model parameters of our Bayesian parameter estimator are shown in Fig--\ref{fig:Best_parameter_distribution}.

\textbf{Case--VI}: In this case,  $30<E_0<45$, $0<a<10$, $-20<b_1<0$, $30<E'_0<45$, $-20<b_2<10$, $T_c=57341$. The assumption is that the line energy depends both on time and flux and there is an explicit rise of  $E_{\text{cyc}}$ with time after $T_c$ = 57341.
This is equivalent to the model described in \citet{2017A&A...606L..13S}. With this model we find the $\chi^2$/d.o.f to be 22/22, with $b_2$=$-8.64^{+4.05}_{-4.15}$. This clearly shows that current observations do not support a recent \textit{turn-up} after 2015.

\textbf{Case--VII}: In this case we assume,   $30<E_0<45$, $0<a<10$, $-20<b_1<0$,
$30<E'_0<45$, $-20<b_2<10$, $50000<T_c<59000$. This is akin to case--VI but here the $T_c$ is found by fitting. In this case we get two degenerate solutions as shown in Fig--\ref{fig:degenerate_parameter_distribution}. The degeneracy can be broken by considering two different ranges of $T_c$; i)  $50000< T_c <
56000$ (VII.a, Fig--\ref{fig:Before_56000_parameter_distribution}) and ii) $57000 < T_c < 59000$ (VII.b, Fig--\ref{fig:after57000_parameter_distribution}). We get two sets of parameter values VII.a and VII.b, (Table--\ref{parameters}) with $\chi^2$/d.o.f of 14.4/21 and 22/21 respectively. The parameter values of VII.b are found to be the same as in case--VI. The parameter values for case--VII.a are found to be close to those of case--V. The value of $b_2$ for VII.a is found to be $(1.04\pm0.86)\times10^{-4}$keV/d, which is close to zero. The chance probability of case--VII.a being identical to case--V is found to be very high (0.18). Hence case--V is considered as the best model because it has fewer parameters. 

\begin{figure*}
\includegraphics[width=17cm]{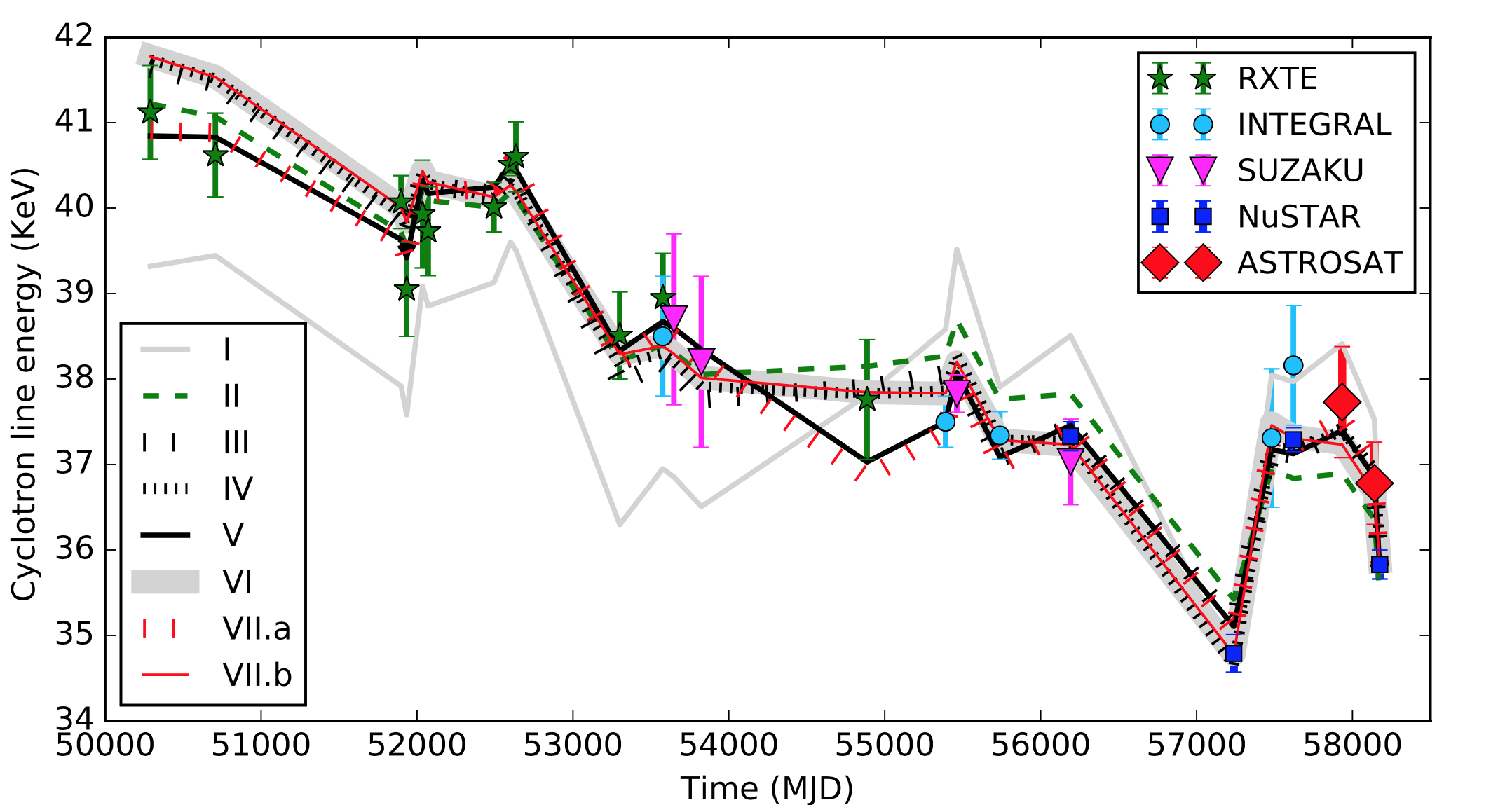}
\caption{The figure shows the variation of the cyclotron line energy ( $E_{\text{cyc}}$) of Her X-1 with time. Data of earlier observations have been taken from various sources as mentioned in \citet{2017A&A...606L..13S}. The flux and the temporal dependence of  $E_{\text{cyc}}$ have been described with Eq--\ref{flux_time_eq} and various combination of parameters. Case--I assumes that no time correlation exists and all the variation of line energy is due to different \textit{Main-On} fluxes. case--II assumes that both flux and temporal correlation co-exist and those are not changing with time. Cases III, IV and V assume that the flux correlation is the same at all times but the line energy has no explicit time dependence after time $T_c$ (MJD 55400, MJD 57341 and MJD $54487^{+515}_{-469}$ respectively). Case--VI assumes that there is flux and temporal dependence but a change in the slope of the temporal correlation occurs after MJD 57341, when the observed line energy exhibited a minimum. case--VII.a and VII.b are the same as case--VI but here $T_c$  is a variable parameter, constrained within;  50000< $T_c$ <
56000 and 57000 < $T_c$ < 59000 respectively.}
\label{fig:energy}
\end{figure*}

So, We find the best fit when we assume that there is an initial decay of line energy with time, and after MJD $54487^{+515}_{-469}$, the line energy no longer has explicit temporal dependence (case--V). In Fig--\ref{fig:energy} we have shown all the data which are considered in our fitting and the behaviour of our model (Eq--\ref{flux_time_eq}) for different cases.  In this correlation analysis a few extreme outlier data points have been ignored (following \citet{2016A&A...590A..91S}) while carrying out the fitting. However the conclusions do not change even if all the observed data are included, albeit with a poorer reduced $\chi^2$. The parameters estimated after including all the data points are given in Appendix--\ref{Tab:relations_app}.

\begin{table*}
\caption{ Values of parameters for describing the variation of the cyclotron line in Her X-1 by Eq--\ref{flux_time_eq}. The flux and temporal dependence have been described for 6 different cases. In case--I, we assume that no time correlation exists and the variation of the line energy is due to different \textit{Main-On} fluxes. case--II assumes that both flux and temporal correlations co-exist and those are not changing with time. Case III,IV and V assume that the flux correlation is the same at all times, but the line energy is constant after time $T_c$ (MJD 55400, MJD 57239 and MJD $54487^{+515}_{-469}$ respectively). Case--VI assumes that the flux correlation is the same at all times and after $T_c$ (57239 MJD, when the cyclotron line energy is found to be minimum) the temporal dependence has changed. CASE VII.a and VII.b are the same as case--VI but here the $T_c$ is found by fitting within the constraint; 50000< $T_c$ <
56000 and 57000 < $T_c$ < 59000 respectively. The F-test values are computed between two cases which are given in Latin numerals under the 'F-values' column.} All errors are reported with 1$\sigma$ (68\%) confidence range.
\begin{center}
\renewcommand{\arraystretch}{1.50}
\begin{tabular}{p{0.65cm}p{1.4cm}p{1.5cm}p{1.4cm}p{1.4cm}p{1.8cm}p{1.4cm}p{1.0cm}p{1.6cm}p{1.4cm}}
\hline 
Case & $E_0$  & $a$ [keV/-  & $b_1$  &   $E'_0$ (keV) & $T_c$  & $b_2$ &  $\chi^2$/d.o.f & F-values & Chance \\
 &  (keV) & ASM-counts/s] & [$10^{-4}$keV/d] &  (keV) & MJD & [$10^{-4}$keV/d]  &  & & Probability \\ 
 \hline
 Range & [35, 45] & [0, 2] & [-20, 0] & [30, 45] & [50000, 59000] & [-20, 10] & & & \\
\hline \hline
 I & $38.72\pm0.06$ &  $1.05\pm0.04$  & - & -& - & - & $346.7/25$ & &\\
 II & $39.30\pm0.07$ &  $0.51\pm0.05$ & $-5.07\pm0.31$ &- & - &- & 77.9/24 & (I--II) 82.8  & $3\times 10^{-9}$ \\
III & $39.30\pm0.11$  & $0.64\pm0.05$ & $-6.15\pm0.83$ & $37.57\pm0.09$ & 55400 & - & 25.8/23 & (II--III) 46.4 & $6\times10^{-7}$ \\
 IV & $39.23\pm0.07$ & $0.52\pm0.05$ & $-6.88\pm0.40$  &$37.55\pm0.13$ &  57341 & - & 26/23 & (II--IV) 45.9 & $6.5\times 10^{-7}$\\
 V & $39.74\pm0.17$ & $0.65\pm0.05$ & $-2.49^{+1.32}_{-1.42}$ & $37.58\pm0.09$ & $54487^{+515}_{-469}$ & - & 15.7/22 & (II--V) 43.6 & $2.2\times 10^{-8}$\\
VI & $39.23\pm0.07$ & $0.47\pm0.06$ & $-7.1\pm0.41$  &$37.90\pm0.21$ &  57341 &  $-8.64^{+4.05}_{-4.15}$ & $22/22$ & - & -\\
VII.a & $39.73^{+0.16}_{-0.17}$ & $0.68\pm0.06$ & $-2.42^{+1.31}_{-1.40}$ & $37.39^{+0.18}_{-0.19}$ & $54565^{+546}_{-515}$ & $1.04^{+0.86}_{-0.85}$ & 14.4/21  & (V-VII.a) 1.9 & 0.18 \\
VII.b & $39.23^{+0.07}_{-0.07}$ & $0.47\pm0.06$ & $-7.08\pm{0.41}$ & $37.84^{+0.23}_{-0.21}$ & $57381^{+105}_{-102}$ & $-8.37^{+4.43}_{-4.27}$ & 22/21  & - & - \\
\hline \hline
\label{Tab:relations}
\end{tabular}
\end{center}
\end{table*}


\section{Discussion and conclusions}
In our work we find that the currently available data do not support the idea of a recent trend of secular increase of the cyclotron line energy  $E_{\text{cyc}}$ or the initiation of a new \textit{Turn-up} phase of Her X-1. The sudden increase of cyclotron line energy with time has been seen in this source 
during 1990-1993 \citep{2001ApJ.Gruber.562..499G}. After that it was believed that a continuous and almost steady decay of cyclotron line energy with time occurred over a period of 20 years at a 
rate of $\sim0.36$ keV yr$^{-1}$ between 2004-2010 \citep{2014A&A...572A.119S}. We have tested different models of the evolution of cyclotron line energy against the available data. We observe that the variation of cyclotron line energy in Her X-1 can be best described by a model that considers an initial decay of line energy after the sudden increase in 1990-1993, and then constancy after a time $T_c$ (Case--V). We determine $T_c$ to be in the range MJD $54487^{+515}_{-469}$, close to that reported in \citet{2019MNRAS.484.3797J}. The relatively large uncertainty in $T_c$ is due to the lack of 
observations in the intervening period. With our best fit model (case--V), the coefficient of flux dependence is found to be $0.65\pm0.05$ keV/ASM-counts/s, close to the value reported by \citet{2007ESASP.622..465S} at the discovery
of the flux dependence (also used by \citet{2019MNRAS.484.3797J}). However the decay rate $-2.49^{+1.32}_{-1.42}\times10^{-4}$ keV/day is found in this case to be significantly different from the values reported earlier \citep{2014arXiv1412.8067S,2016A&A...590A..91S, 2019MNRAS.484.3797J}. This is also seen in one other case (VII.a), among those explored by us.\par

As mentioned in section--1, and derived in the model fits above, Her X-1 exhibits a positive correlation between its luminosity and the cyclotron line energy.  Several explanations have been offered in the literature for this behaviour (e.g. \citet{2007ESASP.622..465S, 2012A&A...544A.123B, 2015MNRAS.454.2714M}). Among the causes cited, the 
most popular is the change in the height of the emission region above the 
neutron  star surface with the accretion rate $\dot{M}$. A reduction in the 
height with increasing $\dot{M}$ would lead to an increase in cyclotron line 
energy \citep{2012A&A...544A.123B}. Other possibilities include a change in   
the velocity distribution of infalling material and a consequent change in   
Doppler effect, as the accretion rate varies \citep{2015MNRAS.454.2714M}. 
\citet{2012MNRAS.420..720M} have shown that larger mass loading in the accretion 
column results in a greater distortion of the magnetic field lines, enhancing 
the field strength at the edge of the polar cap. This would lead to an increase
in the cyclotron line energy, particularly if a fan beam emanating from the 
sides of the accretion column intercepts the observer.

 The secular variation of line energy with time, however, still lacks a clear explanation (e.g. \citet{2014arXiv1412.8067S}).
Possible reasons include a change in the intrinsic dipolar magnetic field strength of the neutron star and changes in the line forming region. The time scale for any 
appreciable  change in the intrinsic dipole moment of the star is expected to be much 
larger ($\sim$million years: \citet{1992A&A.Dipankar..254..198B}, \citet{1994ApJ.Urpin..433..780U}) than the observed time scale of secular variation 
in line energy. An apparent reduction in the local magnetic field may therefore be 
attributed to a screening or burial of the magnetic field at the polar cap by 
continuous accretion \citep{2001ApJ...557..958C}. Such a mechanism would then have to
explain a decay rate of 0.2\% per year in the local magnetic field strength, as 
implied by the observed change in the line energy. In this picture, a sudden increase
of line energy could be due to a re-emergence of the buried field, but the relevant 
time scale for such emergence could be $\sim$100 years \citep{2001ApJ...557..958C, 
1987ApJ...323L..61S} rather than a few years as seen in Her X-1.  
\citet{2017A&A...606L..13S} explain the observed temporal variation of cyclotron line energy  by considering the formation of an accretion mound
at the polar cap region. In the accretion mound, there is a continuous inflow of material as well as an outflow due to the leakage of mater through magnetic 
field lines. If the inflow is greater than the outflow then the mound height increases which leads to the shifting of the line forming region 
away from the surface of neutron star, dragging the central field lines towards the edge.  This reduces the magnetic field intensity of the line forming region, and hence a decay of the cyclotron line energy is expected in case of pencil beam emission \citep{2012MNRAS.420..720M}. When the inflow and outflow of the material balance each other, a dynamic stability arises and the mound height as well as the cyclotron line energy becomes constant. If the accretion mound reaches a critical height, then instabilities in the accretion mound \citep{2013MNRAS.430.1976M, 2013MNRAS.435..718M} may cause a sudden outflow of a substantial amount of matter. The reduction of the mound height and the consequent increase of the central field strength would then cause a sudden increase of the cyclotron line energy. This explanation would, however, contradict the observed positive correlation of line energy with luminosity. To explain the observed luminosity dependence with this model, the source Her X-1 needs to have a fan-beam emission as discussed in the previous paragraph. \par
\begin{figure}
\includegraphics[width=\columnwidth]{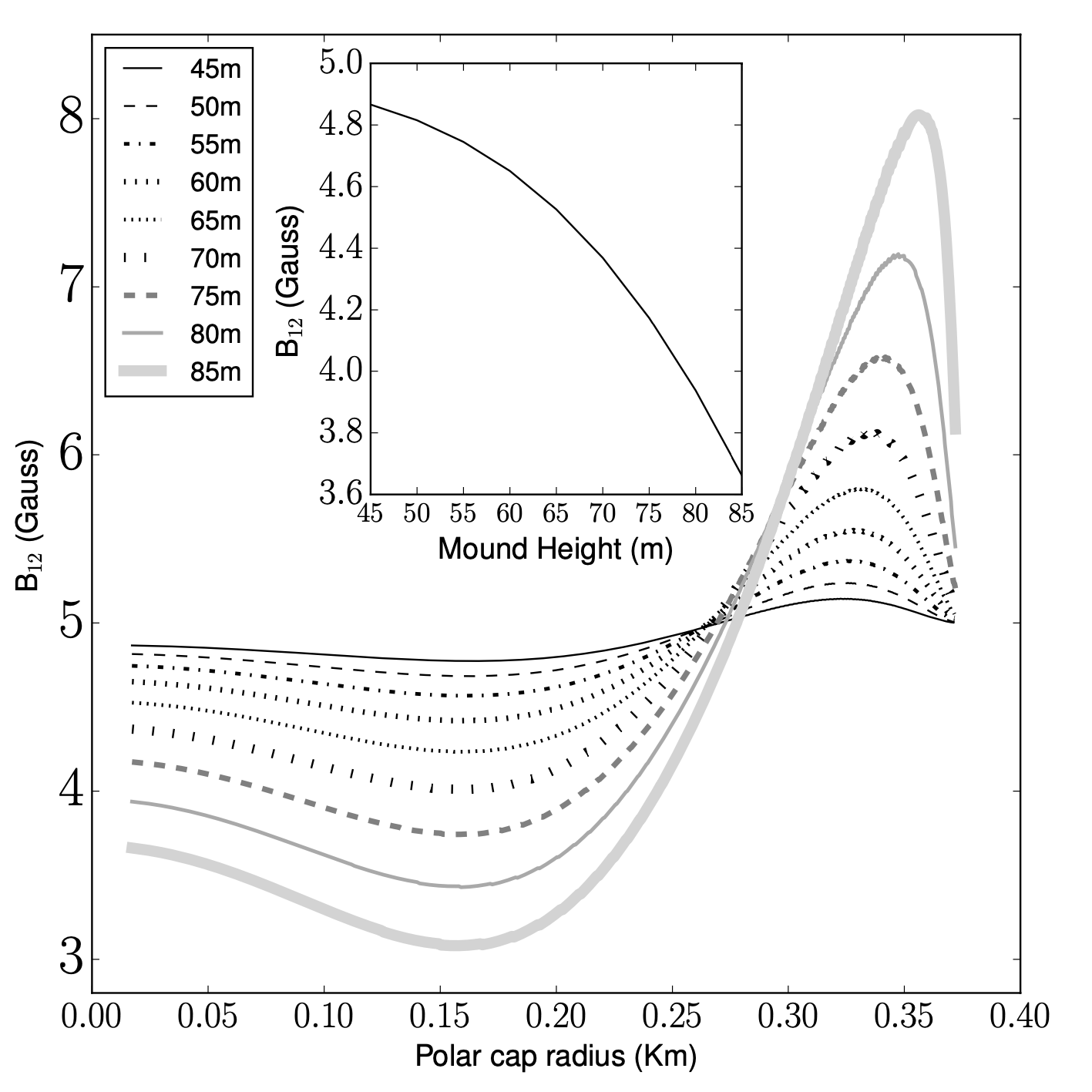}
\caption{Variation of magnetic field strength at the top of the polar cap for different mound heights, for an original dipolar field strength of $\sim5\times10^{12}$ Gauss. The central field strength reduces with the increase of mound height. A reduction of $\sim$8\% in magnetic field strength from the intrinsic value at the pole is observed for a $\sim$60-70 m mound. }
\label{fig:B_R_z}
\end{figure}

We speculate that the positive correlation of the cyclotron line energy with luminosity reflects the change in the height of the emission region above 
the accretion mound \citep{2012A&A...544A.123B}, a larger accretion rate pushing the line forming region close to the mound surface. However, with time there 
is some accumulation of material at the polar cap, generally increasing the
mound height. This is accompanied by a drop in the magnetic field
intensity at the mound center as well as in the line forming 
region above the mound. Once the mound height achieves stability, the secular variation of the line energy would stop. We have extended the work of \citet{2012MNRAS.420..720M} to Her X-1 and show the variation of the magnetic field strength at the top of the mound of different heights in 
Fig--\ref{fig:B_R_z}. We assume the shape of the mound to be parabolic with
an intrinsic dipolar field strength $\sim5\times10^{12}$ G and solve the 
magnetostatic equilibrium for matter confined at the polar cap, as described in \citet{2012MNRAS.420..720M}.  An increase in the height of the mound 
results in more magnetic flux being displaced from the centre towards the mound edge. If the radiation beam is axial and the cyclotron line carries the signature of the magnetic field near the centre of the polar cap, then with a larger mass loading the local field strength will reduce, causing the cyclotron line energy to drop.  In our analysis, an $\sim$8\% decrease in cyclotron line energy over time has been observed, which would be expected from the formation of a mound of height 60-70 m, reducing the central field to ~$4.6\times10^{12}$ G. If a mound had been pre-existing, then the amount of increase in mound height required to cause a similar change may be obtained from the inset in Fig--\ref{fig:B_R_z}.
We find that the  maximum sustainable stable mound height (critical height) for Her X-1 can be $\sim$100 m, beyond which instabilities grow rapidly. Such a scenario may indeed be responsible for the 
observed behaviour of Her X-1. During $\sim$1990-1993, the mound might have reached its critical height, triggering a sudden outflow that reduced the mound height. This would cause an increase in the observed cyclotron line energy. After that the mound height starts to increase and the cyclotron line energy exhibits a 
secular decrease (1995 to 2009: case--V). Then as a dynamic stability between the 
inflow and the outflow of material is established, the mound height as well as the 
cyclotron line energy becomes constant.
It is possible that in future if the inflow into the 
mound exceeds the outflow, then the mound can again 
grow to its critical height, triggering an instability 
in the mound. Regular monitoring of the cyclotron line 
energy of Her X-1 will be required to track such a 
behaviour and to estimate the timescale in which 
further changes in the line energy may occur.

\section*{Acknowledgements}
The author, S. Bala wants to thank Dr Jayashree Roy, Prof. Ranjeev Misra and Prof. Gulab Chand Dewangan, Prof. Amit Shukla, Prof. Biswajit Paul and Dr Niladri Paul for their useful suggestions and discussions. The author would like to thank the anonymous referee for valuable suggestions to improve the work. The research work is funded by the University Grants Commission (UGC). The author wants to thank Inter University Center for Astronomy and Astrophysics (IUCAA) for the research facilities. This publication uses the data from the AstroSat mission of the Indian Space Research Organisation (ISRO), archived at the Indian Space Science Data Centre (ISSDC). This work has used the data from the LAXPC developed at TIFR, Mumbai, and the LAXPC POC at TIFR is thanked for verifying and releasing the data via the ISSDC data archive and providing the necessary software tools. This work has used the data from the Soft X-ray Telescope (SXT) developed at TIFR, Mumbai, and the SXT POC at TIFR is thanked for verifying and releasing the data via the ISSDC data archive and providing the necessary software tools.
 This research has made use of software and/or web tools obtained from the High Energy Astrophysics Science Archive Research Center (HEASARC), a service of the Astrophysics Science Division at NASA/GSFC and of the Smithsonian Astrophysical Observatory's High Energy Astrophysics Division.

\section*{Data availability}
The data underlying this article are publicly available in ISSDC, at https://astrobrowse.issdc.gov.in/astro\_archive/archive/Home.jsp.




\bibliographystyle{mnras}
\bibliography{bala}


\appendix
\section{Background normalisation of LAXPC}
\label{back_earth_occ}
\begin{figure*}
\centering
\includegraphics[width=17cm]{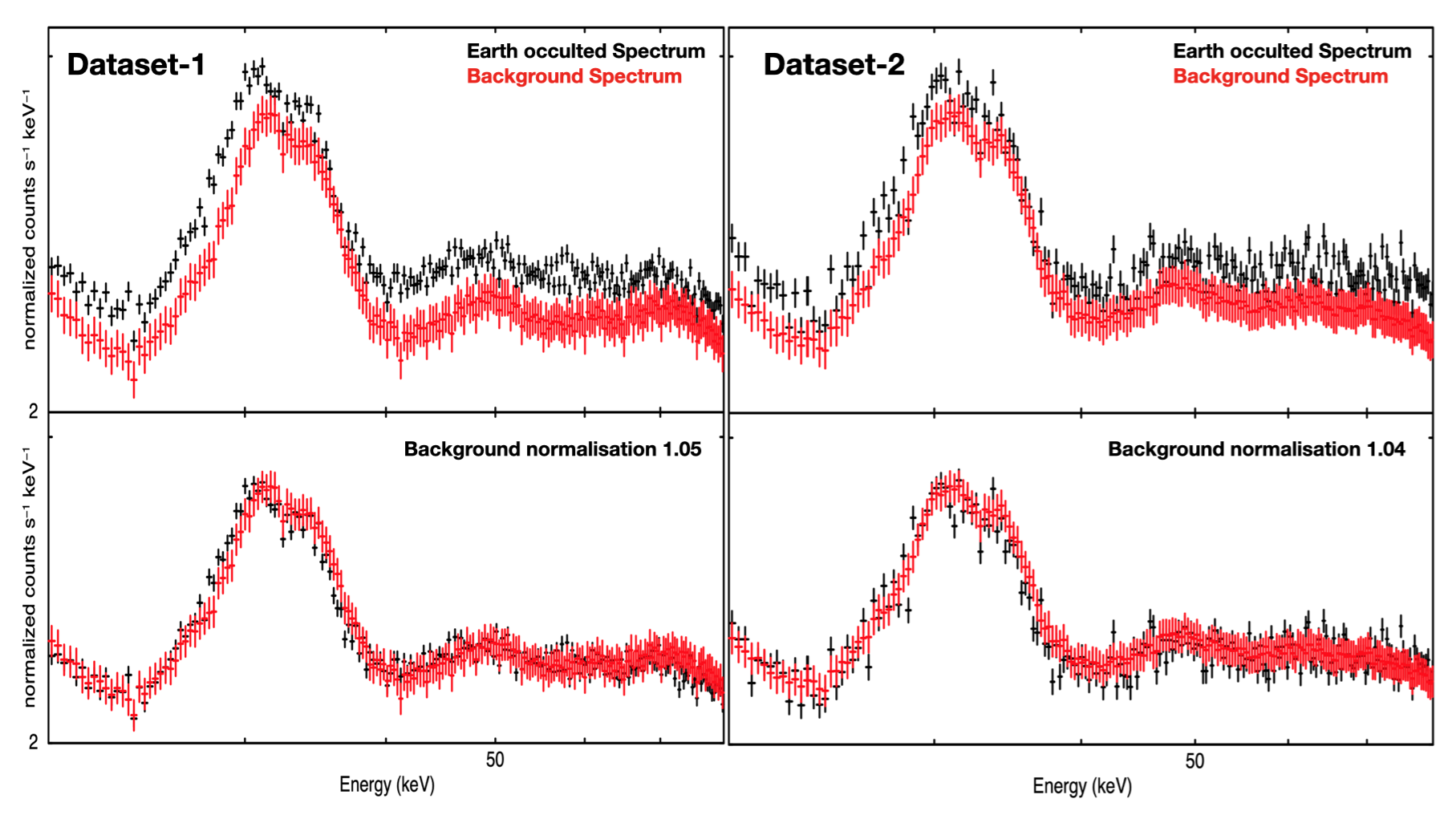}
\caption{Upper panels show the 20-80 keV Earth occulted spectrum (black) and the background spectrum (red) of dataset-1 (left) and dataset-2 (right). Lower panel shows the same, but after using the background rescaling factors 1.05 and 1.04, for dataset-1 (left) and dataset-2 (right) respectively.}
\label{fig:back_earh_occulted}
\end{figure*}

The LAXPC has a known issue of inadequate modelling of the background spectrum. 
 The background model has uncertainties up to 5\% \citep{2017ApJS_antia_LAXPC}.
This mainly effects the high energy end of the spectrum of weak and moderate sources.
The 20.0--60.0 keV source and background count rates of Her X-1 are found to be 69 counts/s and 136.4 counts/s, for 
dataset-1.
The same for dataset-2 is found to be  81.1 counts/s and
140.5 counts/s respectively. 
 In our analysis, we have used different rescaling factors of original background spectra (generated by LAXPC pipeline), to achieve the best fits.
The best fits are found when we rescaled the original background spectra by the factors 1.05 and 1.04 for dataset-1 and dataset-2 respectively.
To crosscheck the justification of the background rescaling factors, we have compared the background spectra with the Earth occulted spectra.
At high energies, the background count-rates (136.4 counts/s for dataset-1 and 140.5 counts/s for dataset-2) are found to be less than the Earth occulted spectrum (143 counts/s for dataset-1 and 145.7 for dataset-2 counts/s), as shown in Fig--\ref{fig:back_earh_occulted}.
We find that the background rescaling values of 1.05 and 1.04 are required to compensate these differences for dataset-1 and dataset-2 respectively.

\section{Creating An ARF of LAXPC}
\label{ARF}
\begin{figure*}
\centering
\includegraphics[width=17cm]{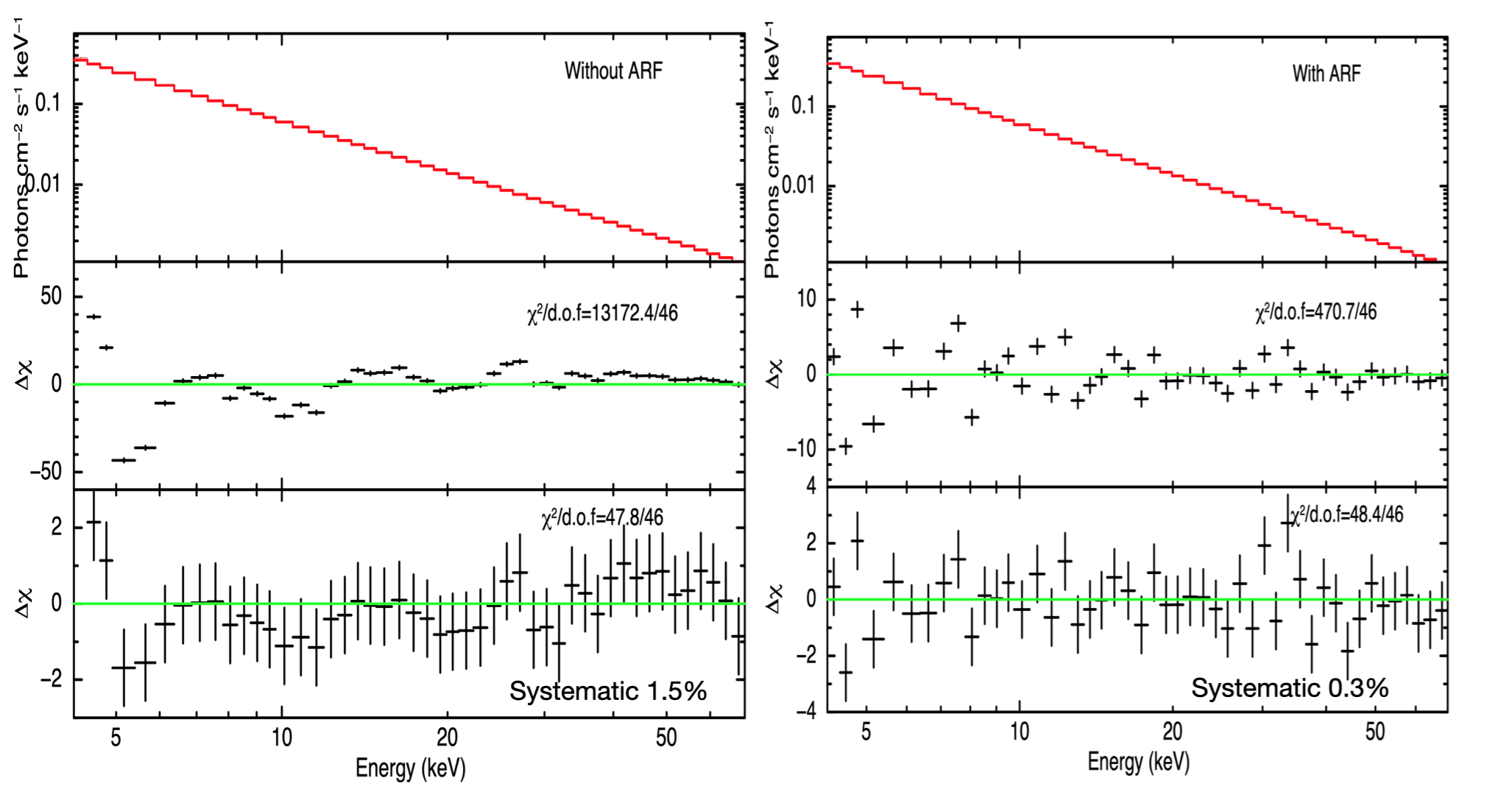}
\caption{4.0-70.0 keV Crab spectrum observed on 16th Jan 2018. Left is the plot of best fit delchi without the ARF and the right is the best fit after applying the ARF. Upper panels show the unfolded Crab spectrum fitted with a power law model. Middle panels show the delchi plots without adding any systematic to the data. Lower panel shows the delchi plots after adding systematic. A systematic of 1.5\% is required to achieve an acceptable reduced chi-square without the ARF and a 0.3\% systematic is required to achieve the same when we have included the ARF.}
\label{fig:crab_no_arf_1}
\end{figure*}

 The estimated systematic error included while fitting the LAXPC spectrum
is usually very large ($\sim$ 2\%-3\%).  With such high
systematics, it is very challenging to find and constrain the presence
of any cyclotron lines which are often detected as weak absorption like features against the continuum in many X-ray spectrum. We observe some unusual dips 
 in the residuals after fitting many LAXPC spectra of different objects. Even for the Crab, which is known to possess a pure power-law
spectrum, these unusual dips have been observed and the locations of the dips were found to be varying with time or the instrument's gain.
This indicates a deficiency in the characterization of the background
and the response. In order to mitigate this effect we
have attempted to create an empirical Auxiliary Response File (ARF)
that models the residuals present in the Crab spectrum.
The Crab
spectrum is a standard power law, with an established photon index
$2.106\pm0.006$, and a low energy absorption component with $N_{\rm H} \sim 4 \times10^{21}$ cm$^{-2}$ \citep{2017ApJ...841...56M}. As the residuals are also found
to be varying with time or the gain of the LAXPC. We chose the closest
possible Crab spectrum (obs date: 16th Jan, 2018, MJD 58134 for dataset-1) to our observation, and modeled the Crab spectrum
after removing all the residuals by the ARF created for it. We found
a major change in the required systematic and the uncertainties of
the parameters after including our ARF.  Fig--\ref{fig:crab_no_arf_1}  shows the best fit model of Crab
spectrum and its residual with and without the ARF.
The requirement of systematic
error in the 4-70 keV band reduced from 1.5--2.5\%  to about 0.3--0.4 \%, to
achieve the same level of reduced chi-squared (Table--\ref{crab_table}). \par

\begin{figure*}
\centering
\includegraphics[width=17cm]{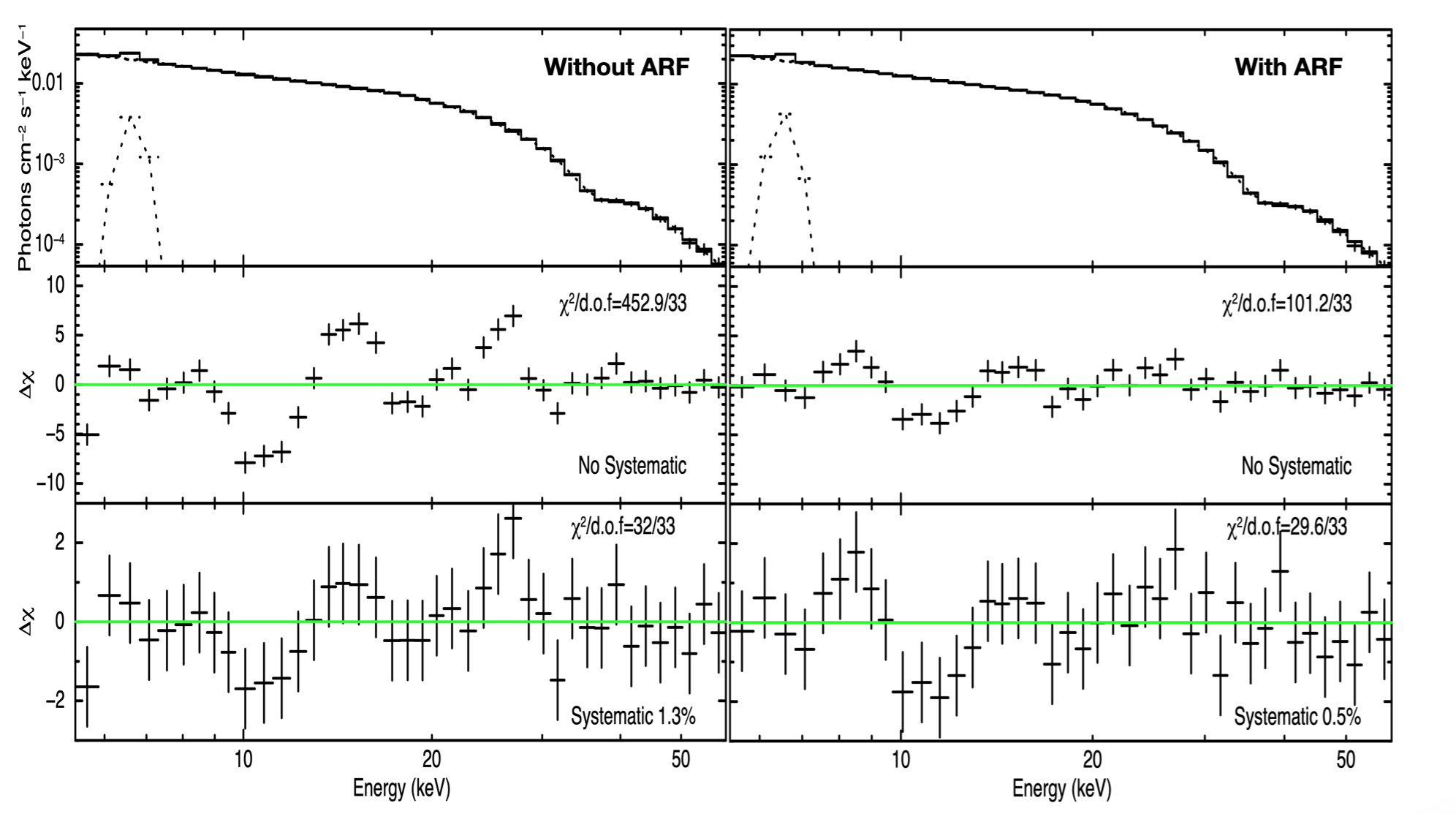}
\caption{Spectrum obtained from dataset-1 of Her X-1 observed by LAXPC. In the figure, top panel shows the unfolded LAXPC spectrum with the fitted model before (left) and after (right) applying our ARF. Middle panel shows the corresponding best fit delchi plot without applying any systematic. The bottom panel shows the best fit delchi plot after applying model systematic (1.3\% without ARF and 0.5\% with ARF) .}
\label{laxpc_only_her_x1}
\end{figure*}

A comparison of the best fit models of only LAXPC data of Her X1, with and without the ARF, has been shown in Table--\ref{parameters_laxpc_only}. The model used to describe the data is,
 \begin{dmath}
\text{I}_{\text{c}}=\text{Constant*(gabs*gabs*highecut*powerlaw)} \label{model_appendix}
\end{dmath}
For dataset-1, a very significant improvement in the fitting has been noticed after applying the ARF ($\chi^2$/d.o.f=452.9/33, without the arf and $\chi^2$/d.o.f=101.2/33, with the arf, Fig--\ref{laxpc_only_her_x1}. In this case also after using the ARF the requirement of systematic
error in the 5-60 keV band reduced from 1.3\%  to about 0.5\%, to
achieve the same level of reduced chi-squared (Table--\ref{parameters_laxpc_only}).

\begin{table}
\caption{The Best fit model parameter values of Crab spectrum before and after applying our ARF. Errors are reported with 3$\sigma$ (90\%) confidence range.}
\centering
\renewcommand{\arraystretch}{1.50}
\begin{tabular}{|c|c|c|}
\hline

Parameter & Without ARF & With ARF \\
\hline 
\hline
$N_{\text{H}}$ ($10^{22}$cm$^{-2}$) & $0.4^*$ & $0.4^*$ \\
$\Gamma$ & $2.09^{+0.01}_{-0.01}$ & $2.108^{+0.001}_{-0.001}$ \\
$norm$ & $7.62^{+0.1}_{-0.1}$ & $7.74^{+0.03}_{-0.03}$  \\
\hline
Systematic & $1.5\%$ & $0.3$\% \\
\hline
$\chi^2$/d.o.f & $47.8/46$ & $48.4/46$  \\
\hline

\hline
\end{tabular}

\label{crab_table}
*The parameter value is fixed.
\end{table}

\begin{center}
\begin{table}
\caption{The best fit model parameters of Her X-1, only using the LAXPC data, with and without including the arf. All the errors are reported with 3$\sigma$ (90\%) confidence range.}
\renewcommand{\arraystretch}{1.60}
\begin{tabular}{|p{2.4cm}|p{2.23cm}|p{2.23cm}|}
\hline
Parameter & With ARF & Without ARF  \\
\hline
$\Gamma$ &               $0.99^*$ & $0.99^*$   \\
 $E_{\text{cut}}$ (keV) &         $20.37^*$ & $20.37^*$  \\
 $E_{\text{fold}}$ (keV) &        $10.01^*$ &  $10.01^*$ \\
$E(K_{\alpha})$ (keV) &   $6.57^{+0.07}_{-0.07}$ & $6.65^{+0.12}_{-0.12}$  \\
$\sigma(K_{\alpha})$ (keV) & $0.22^*$  & $0.22^*$    \\
 $E_{\text{cyc}}$ (keV) &          $36.8^{+0.4}_{-0.37}$ &  $36.47^{+0.4}_{-0.37}$ \\
$\sigma _{\text{cyc}}$ (keV) &     $3.17^{+0.3}_{-0.29}$ &  $2.94^{+0.32}_{-0.31}$ \\
$d _{\text{cyc}}$ &             $5.3^{+0.41}_{-0.38}$ &   $4.78^{+0.4}_{-0.39}$  \\
\hline
$\chi^2$/d.o.f &          $29.6/33$ & $32/33$  \\
\hline
LAXPC Systemnatic &     $0.5\%$ & $1.3\%$  \\
\hline \hline

\end{tabular}

*The parameter value is fixed.\\
Optical depth, $\tau=d/\sqrt{\sigma2\pi}$, All the uncertainties are measured in 90\% confidence level.

\label{parameters_laxpc_only}
\end{table}
\end{center}

\begin{table*}
\caption{Same as Table--\ref{Tab:relations} but including all the observations which have been ignored in \citet{2007ESASP.622..465S}.}
\centering
\renewcommand{\arraystretch}{1.30}
\begin{tabular}{p{0.65cm}p{1.4cm}p{1.5cm}p{1.4cm}p{1.4cm}p{1.8cm}p{1.4cm}p{1.0cm}p{1.6cm}p{1.4cm}}
\hline
Case & $E_0$  & $a$ [keV/-  & $b_1$  &   $E'_0$ & $T_c$  & $b_2$ &  $\chi^2$/d.o.f & F-values & Chance \\
 &  (keV) & ASM-counts/s] & [$10^{-4}$keV/d] &  (keV) & MJD & [$10^{-4}$keV/d]  &  & & Probability \\ 
 \hline
 Range & [35, 45] & [0, 2] & [-20, 0] & [30, 45] & [50000, 59000] & [-20, 10] & & & \\
 
\hline \hline
 I & $38.77\pm0.06$ &  $1\pm0.04$  & - & -& - &-& 412.1/30 & &\\
 II & $39.31\pm0.07$ &  $0.51\pm0.05$ & $-4.98\pm0.28$ &- & - &- & $108.1/29$ & (I--II) 81.6 & $6\times10^{-10}$ \\
III & $39.27\pm0.10$  & $0.63\pm0.05$ & $-6.14\pm0.76$ & $37.66\pm0.09$ & 55400 & - & 74.4/28 & (II--III) 12.7 & $1.3\times10^{-3}$ \\
IV & $39.25\pm0.07$ & $0.52\pm0.05$ & $-6.4\pm0.37$  & $37.55\pm0.12$ &  57341 & - & 71.1/28 & (II--IV) 14.6 & $6.8\times10^{-4}$\\
V & $39.73\pm0.17$ & $0.62\pm0.05$ & $-2.69\pm1.25$ & $37.65\pm0.09$ &  $54093\pm208$ & - & 61.7/27 & (II--V) 10.2 & $5.2\times10^{-4}$ \\
VI & $39.24\pm0.07$ & $0.48\pm0.05$ & $-6.58\pm0.37$  & $37.91\pm0.22$ &  57341 &  $-8.6\pm4.11$ & $67.1/27$ & -  & - \\
VII.a & $39.74\pm0.17$ & $0.6\pm0.05$ & $-2.69\pm1.28$ & $37.78\pm0.20$ & $54081\pm212$ & $-0.57\pm0.76$ & 61.2/26  & (V--VII) 0.2 & 0.65 \\
VII.b & $39.25\pm0.07$ & $0.48\pm0.05$ & $-6.55\pm0.38$ & $37.84\pm0.22$ & $57394\pm126$ & $-8.21\pm4.20$ & 67.2/26  & - & - \\
\hline
\label{Tab:relations_app}
\end{tabular}
\end{table*}


\bsp	
\label{lastpage}
\end{document}